\documentclass[fleqn,usenatbib]{mnras}

\usepackage{newtxtext,newtxmath}

\usepackage[T1]{fontenc}

\DeclareRobustCommand{\VAN}[3]{#2}
\let\VANthebibliography\thebibliography
\def\thebibliography{\DeclareRobustCommand{\VAN}[3]{##3}\VANthebibliography}

\usepackage{graphicx}	
\usepackage{amsmath}	

\newcommand* {\kms}  {\mbox{km\,s$^{-1}$}}
\newcommand* {\Msun} {\mbox{M$_\odot$}}

\title[Spirals and the Cepheid gap]
{Spiral arms and the angular momentum gap in Milky Way Cepheids}

\author[Semczuk et al.]{%
Marcin Semczuk,$^{\!\!1}$
Walter Dehnen,$^{\!\!2,1}$
Ralph Sch{\"o}nrich,$^{\!\!3}$ and
E. Athanassoula,$^{\!\!4}$
\smallskip
\\
$^1$ School for Physics and Astronomy, University of Leicester, University Road, LE1 7RH, UK \\
$^2$ Astronomisches Recheninstitut, Zentrum f{\"u}r Astronomie der Universit{\"a}t Heidelberg, M{\"o}nchhofstra\ss{}e 12-14, 69120, Heidelberg, Germany\\
$^3$ Mullard Space Science Laboratory, University College London, Holmbury St.~Mary, Dorking, Surrey, RH5 6NT, UK \\
$^4$ Aix Marseille Université, CNRS, CNES, LAM, Marseille, France
}

\date{Accepted XXX. Received YYY; in original form ZZZ}

\pubyear{2022}

\begin{document}
\defcitealias{Gaia22}{G22}
\defcitealias{D22}{D22}

\label{firstpage}
\pagerange{\pageref{firstpage}--\pageref{lastpage}}
\maketitle

\begin{abstract}
The angular-momentum distribution of classical Cepheids in the outer Milky Way disc is bi-modal with a gap at $L_\mathrm{gap}=2950\,\kms$kpc, corresponding to $R=13\,$kpc, while no similar feature has been found in the general population of disc stars. We show that star formation in multiple spiral arm segments at the same azimuth leads to such multi-modality which quickly dissolves and only shows in young stars. Unlike other explanations, such as a 1:1 orbital resonance with the Galactic bar, this also accounts for the observed steepening of the stellar warp at $L_\mathrm{gap}$, since the adjacent spiral arms represent different parts of the warped gas disc. In this scenario the gap is clearly present only in young stars, as observed, while most purely stellar dynamical origins would affect all disc populations, including older disc stars.

\end{abstract}

\begin{keywords}
Galaxy: kinematics and dynamics -- Galaxy: structure -- Galaxy: disc -- galaxies: spiral -- stars: variables: Cepheids
\end{keywords}



\section{Introduction}
Classical Cepheids (CCs) are pulsating stars and their well-defined period-luminosity relation makes them excellent distance indicators. With ages of the order of few hundred Myr, they are also quite young. These two characteristics make them ideal tracers of the structure of the young Galactic disc. For example, CCs have been used to map the Galactic stellar warp \citep{Chen2019,Skowron2019} and to estimate its parameters \citep{Lemasle2022}.

For the spiral arms of the Milky Way (MW) the results are certainly less clear than for the warp. This can be said about all spiral tracers in the Galaxy. The main problem is our location within the disc, which limits the view to the far side and hampers our understanding of the global picture. Nevertheless, several authors \citep[e.g.][]{Majaess2009, Veselova2020, Poggio2021, Lemasle2022}  attempted to trace spiral arms with CCs and pointed to the coincidence of their spatial distribution with spiral arms defined in the literature by other tracers. The ages of CCs imply that they could have undergone up to one Galactic rotation near the Solar circle since their formation, which limits their distance from the spiral arm in which they were born. This was noted by \cite{Skowron2019} who discussed a few scenarios in which the current distribution of CCs could have originated from 1, 2 or 3 separate interstellar medium (ISM) spiral arms.

The recent Gaia Data Release 3 of the \citet[][hereafter \citetalias{Gaia22}]{Gaia22} combined proper motions from Gaia EDR3 \citep{Lindegren21} with line-of-sight velocities and distances from the CC period-luminosity relation of \cite{Ripepi2019,Ripepi22} to obtain the 3D kinematics of MW CCs. Using these data, \citet[][hereafter \citetalias{D22}]{D22} analysed the radial dependencies of the azimuthal velocity $v_{\phi}$ and the $z$ component $L_z=Rv_\phi$ of angular momentum for the CCs younger than 200 Myr. They found that these quantities follow a tight relation with radius $R$, showing a gap or local minimum at $L_z\approx2950\,\kms$kpc, see Fig.~1 of \citetalias{D22} or the bottom panel of Fig.~\ref{fig:gap1} below.

\citetalias{D22} discussed several possible explanations of this gap, focusing mostly on orbital resonances with the Galactic bar. Instead in this paper we suggest an alternative and much simpler explanation, namely that the observed gap in angular momentum originates from a spatial gap between two thin spiral arms in the ISM where the CCs formed. We illustrate this idea with a simulated spiral galaxy.

This paper is structured as follows. In Section~\ref{sec:simgal}, we introduce the simulated galaxy and discuss its properties. In Section~\ref{sec:gap}, we describe the phase space gap found in the simulated galaxy and discuss its origin and its connection to the warp. Sections~\ref{sec:discuss} and~\ref{sec:sum} discuss and summarize our findings, respectively.

\begin{figure*}
	\includegraphics[width=16cm]{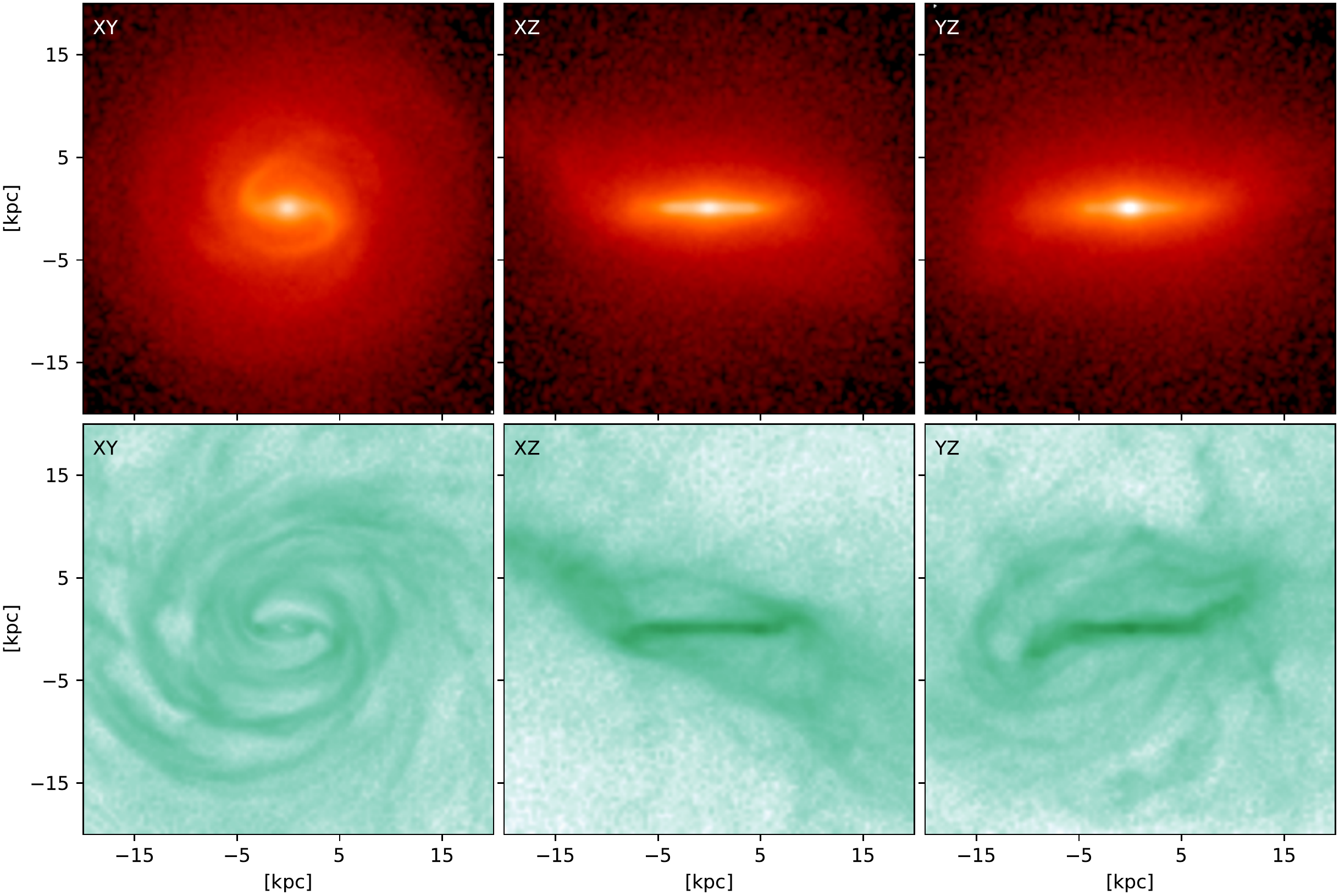}
    \caption{Density distribution of all stellar (top) and gas (bottom) particles of the simulated galaxy in 3 projections as indicated at redshift z=0 ($t=13.803$\,Gyr).}
    \label{fig:map0}
\end{figure*}

\section{The simulated galaxy}
\label{sec:simgal}
We found a simulated galaxy with a similar gap in run TNG50-1 of the magnetohydrodynamical cosmological suite IllustrisTNG \citep{Pillepich2019, Nelson2019}. Baryonic (gas/star) particles in TNG50-1 have a typical mass of $8.5\times10^4\,\Msun$ and a gravitational softening of 288\,pc, implying a resolution sufficiently high to model galactic structures, such as spiral arms, with some detail. Within its simulation volume of (50\,Mpc)$^3$, TNG50-1 has about 130 galaxies as massive as the Milky Way (\citeauthor{Pillepich2019}).

Our example galaxy (ID\,=\,546474 at redshift $z$\,=\,0) is a barred spiral with a total mass of $5.7\times10^{11}\,\Msun$ within 100\,kpc. Inside 20\,kpc the stellar and gas mass are $4.6\times10^{10}\,\Msun$ and $1.4\times10^{10}\,\Msun$, respectively, at $z=0$. Most of the former are in an exponential disc with a scale length of 2.7\,kpc. About 1\,Gyr before the end of the simulation, a central bar formed and reached a radial extent of 3.3\,kpc at $z=0$. This bar rotates with pattern speed 34\,\kms\,kpc$^{-1}$ \citep*[obtained via the unbiased Fourier method of][]{Dehnen22}, which places co-rotation at 6.7\,kpc, about twice the bar radius, implying a slow bar.

Outside the bar region, the stellar and gaseous surface density maps of Fig.~\ref{fig:map0} show $m=2$ spiral arms, presumably due to tidal interactions. In the stellar disc the spirals are much shorter and less pronounced than in the gas, where they are easily traceable to $\sim3$ times larger radii. The edge-on views show a strong warp, which again is more pronounced in the gas. This warp predates the bar, as it can be seen in the gas as early as 3\,Gyr before the end of the simulation (not shown), and likely originates from misaligned gas accretion. More details on the evolution of the simulated galaxy can be found in Appendix A.

\section{The angular momentum gap in simulations}
\label{sec:gap}
\subsection{Gap characterization}

\begin{figure}
	\includegraphics[width=\columnwidth]{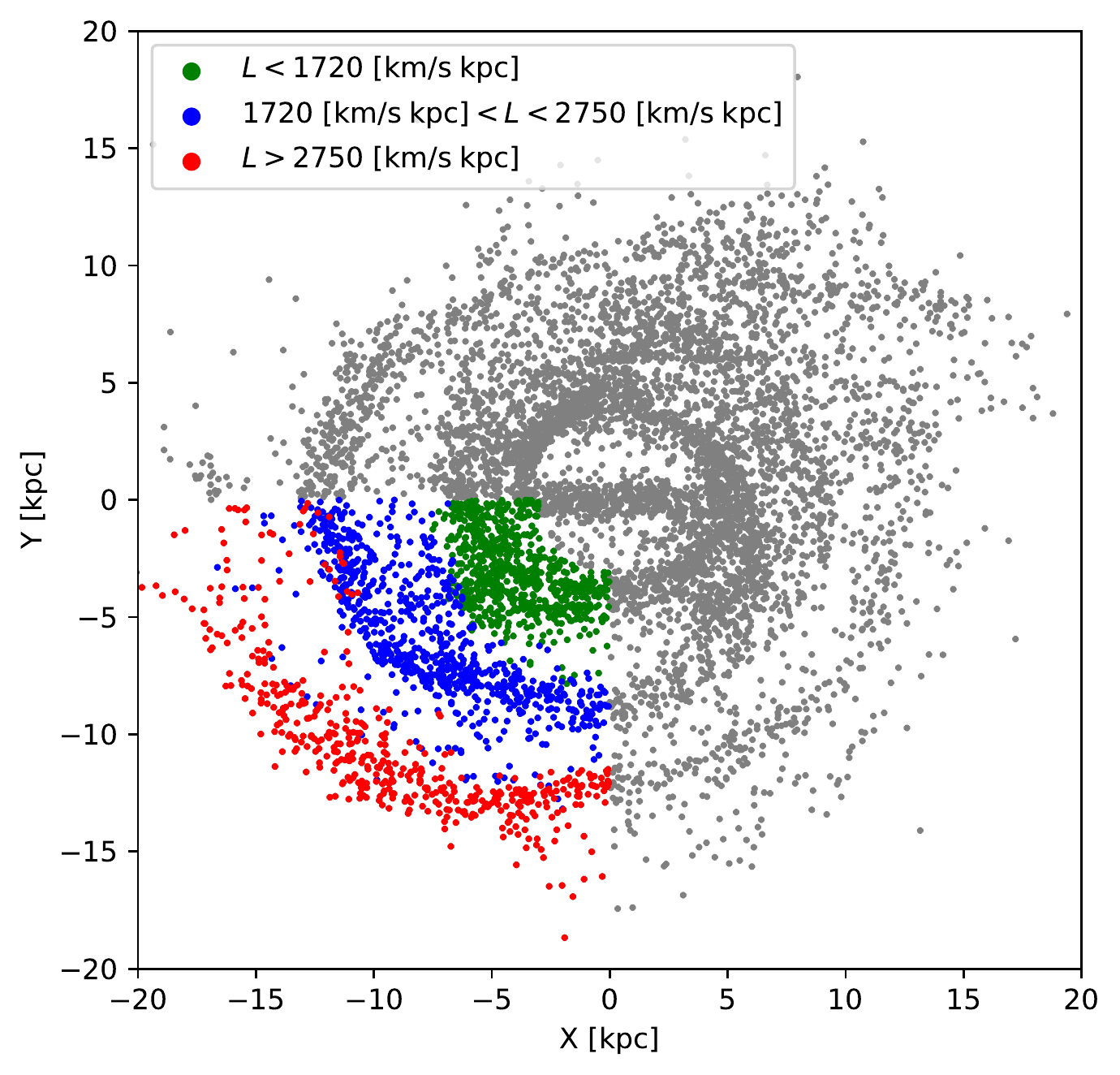}
    \vspace*{-4mm}
    \caption{Spatial distribution of the stellar particles younger than 200 Myr from the simulated galaxy. Colour coded are 3 sub-samples separated by their angular momentum in the quadrant Q3 (see the third row of Fig.~\ref{fig:gap1}).}
    \label{fig:arms0}
\end{figure}

In order to emulate Galactic Cepheids, we select from the simulated galaxy star particles younger than 200\,Myr, the same age limit as that used for CCs by \citetalias{D22} and \citetalias{Gaia22}. The face-on distribution of these particles, shown in Fig.~\ref{fig:arms0}, resembles more the gas morphology in Fig.~\ref{fig:map0} than the overall stellar distribution. Besides the bar and the spirals, these young stars also form a ring of the same size as the bar, i.e. a feature often observed in external galaxies and called an inner ring. 

\begin{figure*}
	\includegraphics[width=17.5cm]{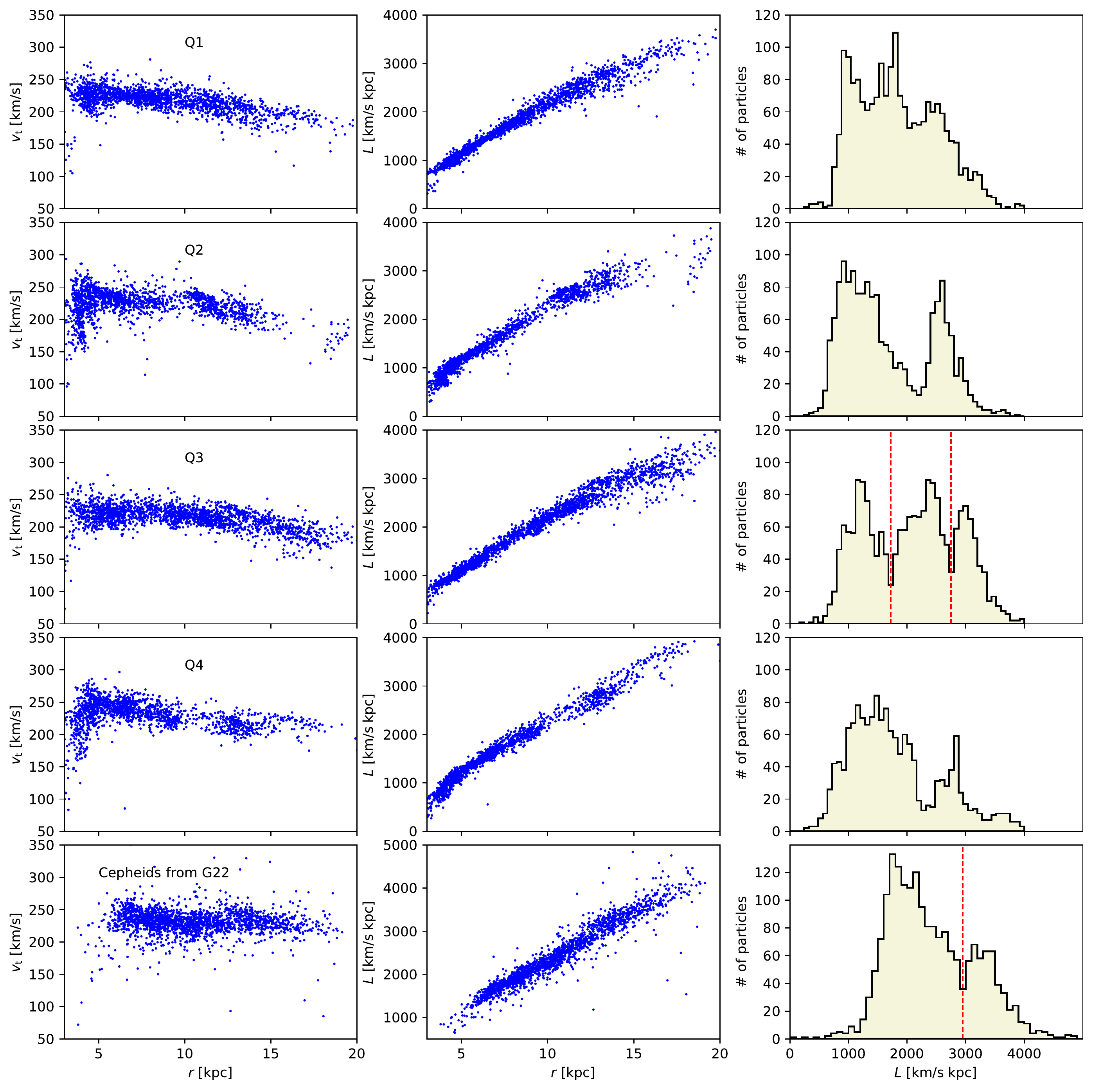}
    \vspace*{-3mm}
    \caption{
    The distributions over spherical radius $r$ and tangential velocity $v_\mathrm{t}$ (left) and total angular momentum $L=rv_\mathrm{t}$ (middle), as well as histograms over $L$ (right) for star particles at $r>3\,$kpc and younger than 200 Myr from each quadrant (as indicated) of the simulated galaxy and for the MW CCs (bottom) provided by \citetalias{Gaia22,D22}.
    }
    \label{fig:gap1}
\end{figure*}

We divide these star particles by their azimuth into four quadrants in a reference frame aligned with the bar (other random orientations do not alter our results significantly; see Fig.~\ref{fig:arms0}) and do not consider particles from the inner 3\,kpc, since \citetalias{Gaia22} do not cover this inner region. In the first four rows of Fig.~\ref{fig:gap1}, we show for each quadrant the radial dependencies of the tangential velocity $v_{\mathrm{t}}$ and the absolute value of the angular momentum $L$ together with histograms of $L$ for these young star particles. In the bottom row of Fig.~\ref{fig:gap1}, we show the same quantities for the MW CCs obtained from \citetalias{Gaia22,D22}\footnote{Deviating from \citetalias{Gaia22}, we used a
Solar velocity of $(13,250,6.9)\kms$ to translate to the Galactic frame, a blend of the results derived by \cite{SBD2010} and \cite{Schoenrich2012}.}. We chose spherical radius $r$, $v_{\mathrm{t}}$, and total angular momentum $L$ here, instead of cylindrical radius $R$, $v_\phi$, and $L_z$ (as \citetalias{D22} did), in order to present a consistent comparison with the simulation. As discussed in Section~\ref{sec:simgal}, the disc of the simulated galaxy is strongly warped when $v_\phi$ and $L_z$ capture the rotational motions only partly, while $v_{\mathrm{t}}$ and $L$ provide the full picture. This change of variables hardly makes a difference for the distribution of MW CCs, as shown in Fig.~1 of \citetalias{D22}. 

Angular momentum $L$ shows tight relations with radius in each quadrant, slightly tighter in fact than MW CCs (which may be a consequence of distance errors). In all quadrants, the distribution has one or two gaps, most notable as dips in the histograms over $L$. Of all quadrants, Q3 best resembles the distributions of Galactic CCs, except that it has not one but two clear gaps, the inner of which at $L=1720\,\kms$kpc has no analogon in the MW CC data.

\begin{figure}
	\includegraphics[width=\columnwidth]{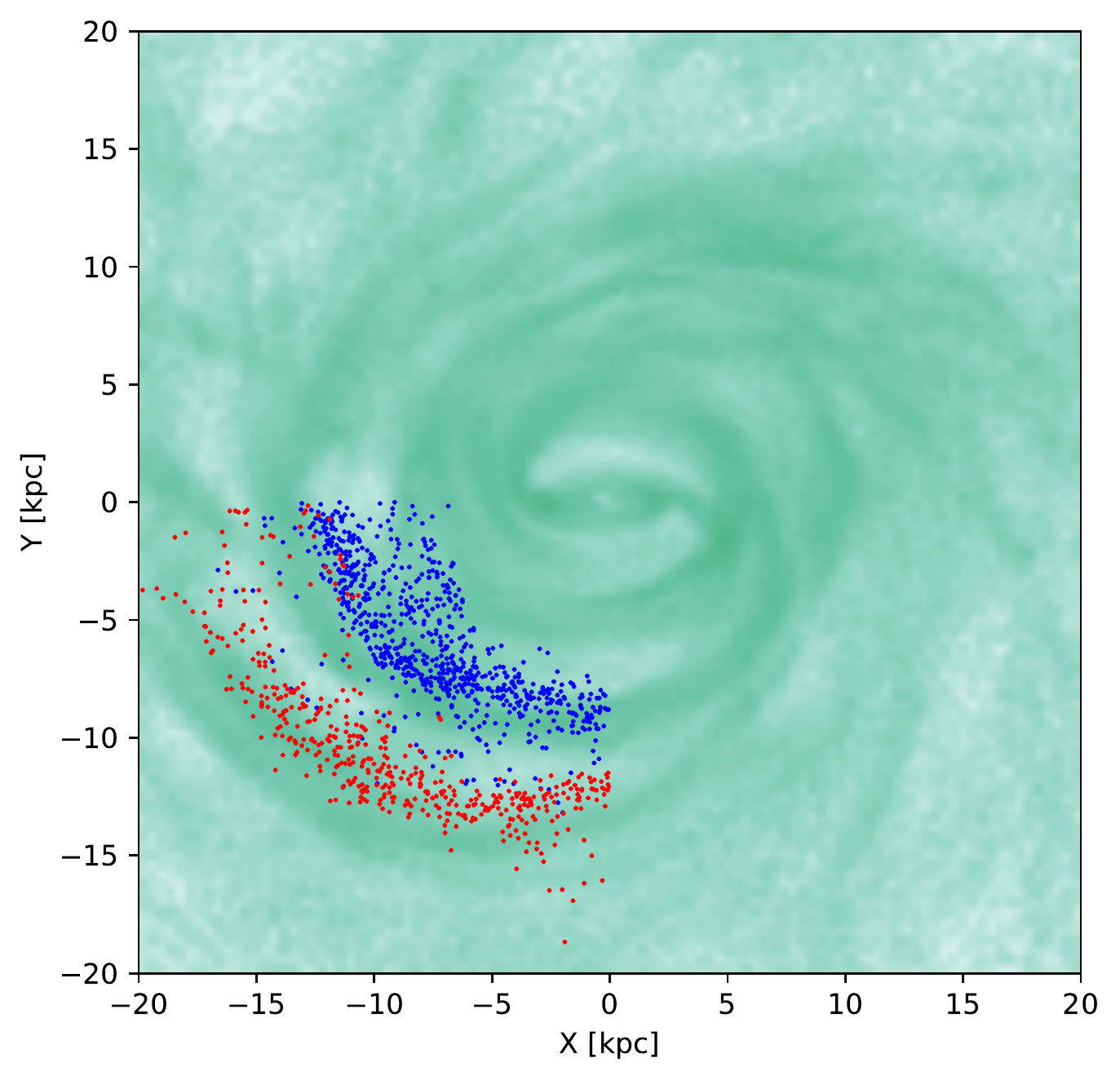}
    \vspace*{-5mm}
    \caption{Face-on density distribution of the gas from the simulated galaxy with overlaid the scatter plot of the young stellar particles from the two spiral arms between which the angular momentum gap has formed. The selection of these particles is in Fig.~\ref{fig:arms0} and~\ref{fig:gap1}.}
    \label{fig:arms1}
\end{figure}

We divide the young stellar particles from Q3 into three sub-samples based only on their angular momentum $L$ (as marked by dotted lines in Fig.~\ref{fig:gap1}) and paint them with different colour in the spatial map of Fig.~\ref{fig:arms0}. From this figure, it is clear that angular momentum separates stars into two spiral arms at different radii and an inner ring near the bar co-rotation radius. The obvious interpretation is that the gaps in angular momentum originate from the spatial gaps between the spiral arms in which these particles actually formed. In Fig.~\ref{fig:arms1} we over-plot the young star particles associated with these two spirals on the gas density map viewed face-on. The star particles lie close to the gaseous spiral arms but slightly lag behind. This is expected because outside its co-rotation radius the spiral propagates faster than circular orbits and hence any stars (some evidence for this to also happen in MW was claimed by \citealt{Hou2015}). Unfortunately the cadence of the TNG50 output data is not sufficient to investigate this in greater detail.

\subsection{Vertical structure}
\citetalias{D22} noted that the radial position of the gap in CCs coincides with the onset of the signal in vertical velocities $v_z$ associated with the Galactic warp. If the gap is caused by some orbital resonance of the in-plane motions (such as an outer 1:1 resonance with the Galactic bar as favoured by \citetalias{D22}), then such a coincidence is unexpected and indeed \citetalias{D22} concluded that its origin is unclear.

\begin{figure}
	\includegraphics[width=\columnwidth]{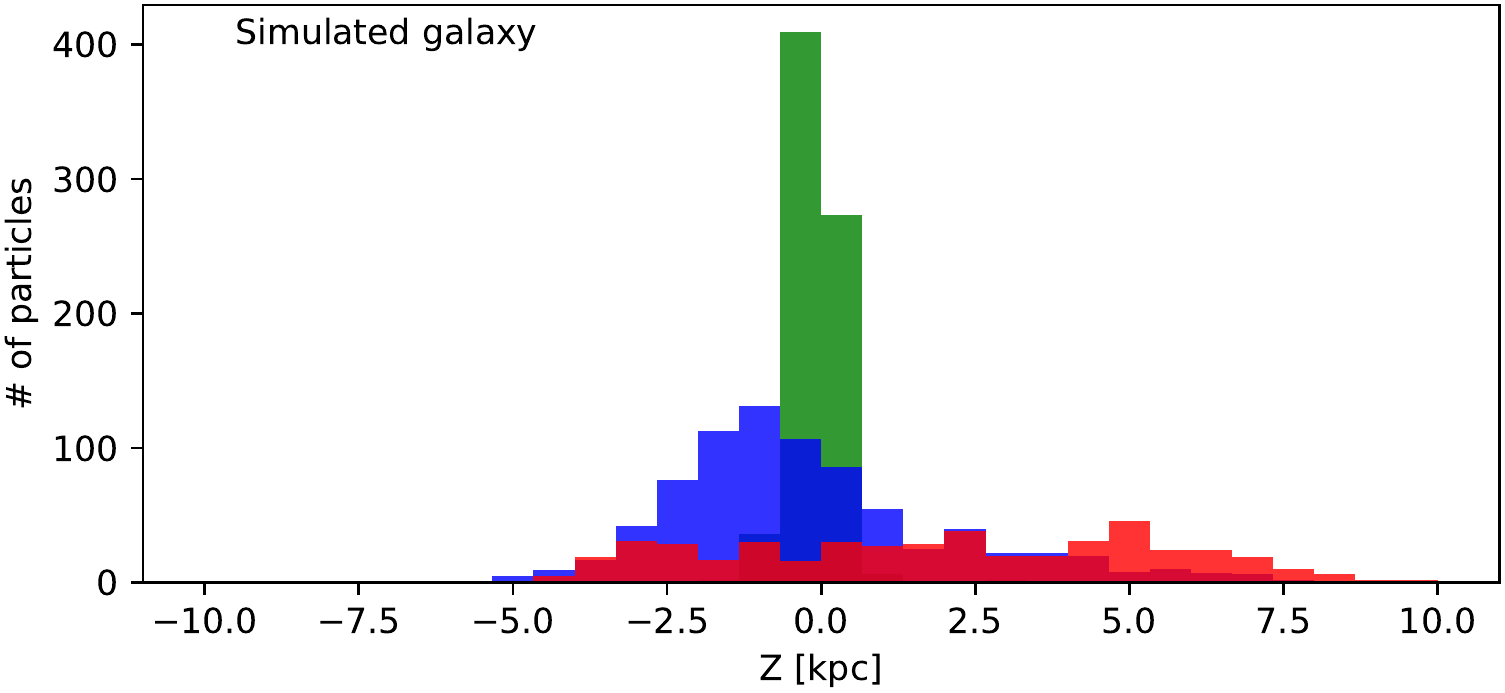}
	\includegraphics[width=\columnwidth]{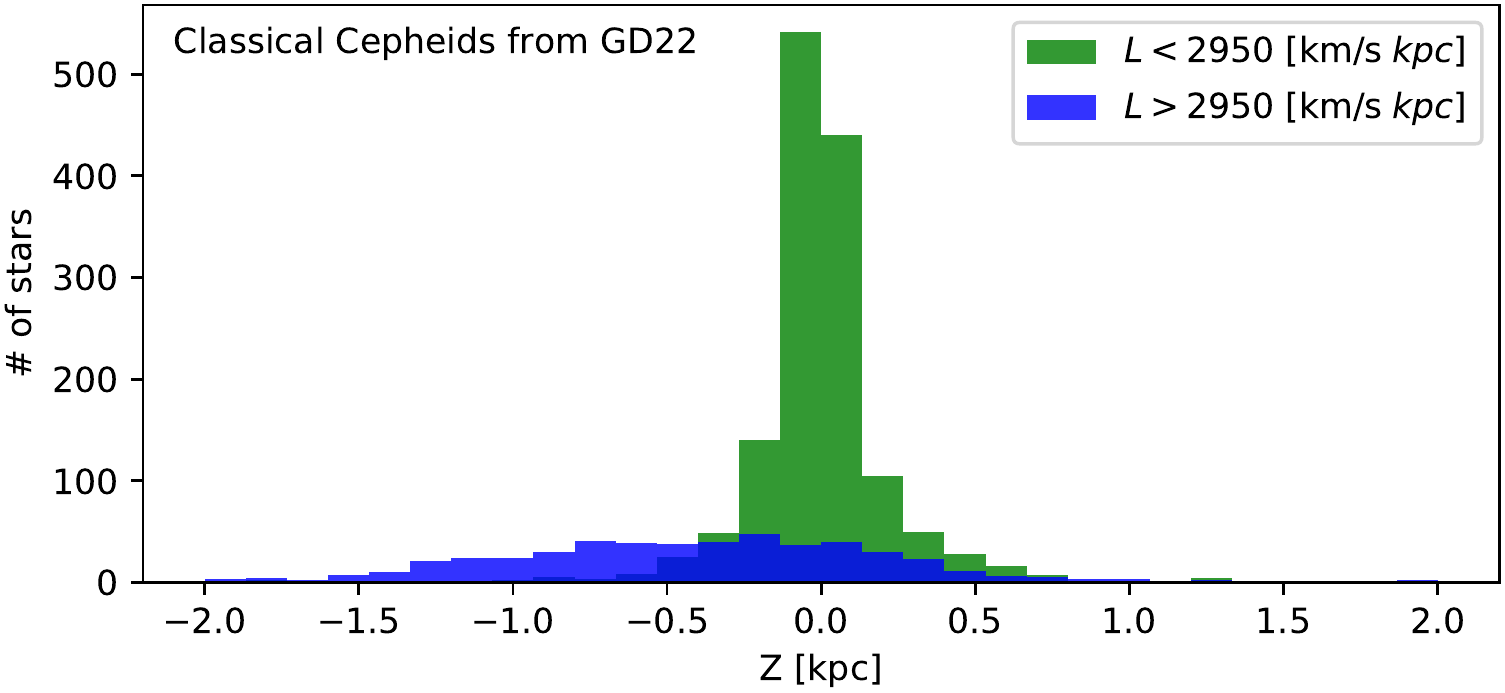}
    \vspace*{-5mm}
    \caption{The distributions of the $z$ coordinates for stellar particles from the simulated galaxy (top) and for Galactic classical Cepheids from the sample of \citetalias{D22} (bottom) for different subsets according to their total angular momentum $L$. The colour coding for the simulated galaxy is the same as in previous figures.
    \label{fig:z_hist}
    }
\end{figure}

We investigate in Fig.~\ref{fig:z_hist} the vertical distributions of the groups of MW (bottom) and simulated (top) CCs separated by the angular momentum gaps. In both cases, the smallest $L$ group (green) is vertically the narrowest and well confined to the galactic midplane, while the groups beyond the respective gaps have very different vertical distributions: much broader and asymmetric.

\begin{figure}
	\includegraphics[width=\columnwidth]{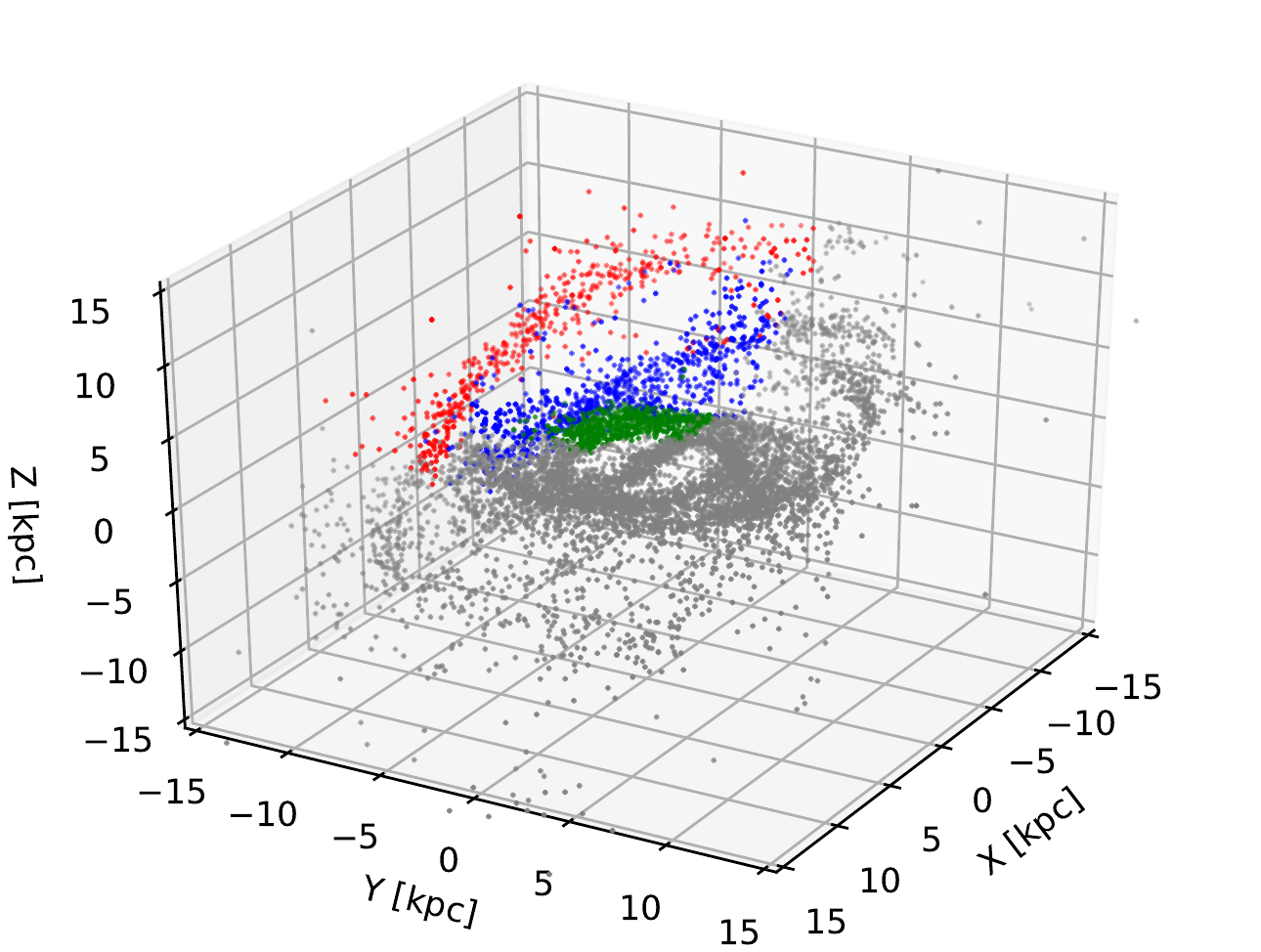}
    \vspace*{-4mm}
    \caption{The 3D projection of the stellar particles younger than 200 Myr from the simulated galaxy. Colour coded are 3 subsamples separated by their angular momentum in the quadrant Q3 (see the third row of Fig.~\ref{fig:gap1}). The projection demonstrates the origin of the vertical separation between the warped spiral arms.  
    \label{fig:3D}
    }
\end{figure}

For the simulated galaxy, Fig.~\ref{fig:3D} presents a 3D projection of the disc of young stars. Quite obviously, the outer spiral arm (red) is strongly warped with respect to the inner ring (green), which is almost aligned with the galactic plane, while the intermediate arm (blue) is intermediately warped between these. So with increasing $L$, the disc of young stars is increasingly warped, reflecting the state of the gas disc at the time of their birth. The same explanation likely holds for the MW CCs, albeit with a much weaker warp: the outer spiral beyond the gap was (at the time of forming the CCs) more inclined to the inner Galactic disc and gave birth to stars on correspondingly inclined orbits.

\section{Discussion}
\label{sec:discuss}
\citetalias{D22} reviews several possible explanations of the CC angular-momentum gap. They dismiss a spiral-arm origin since the gap feature is present in not only the quadrant in which \cite{Poggio2021} identified the main spiral arm. Tracking spiral arms of the MW is traditionally a very difficult task, but we think that the origin from two adjacent spirals at the same azimuth can work even if these arms do not fully cover the same range of azimuths. For example in Fig.~\ref{fig:arms0} the outermost arm (red) could end half way of its current azimuth range and the phase space would still show the gap. The only necessary condition for this mechanism to create a gap is to have two star forming spiral arms in some azimuthal range.  

\citetalias{D22} also consider a scenario in which the interaction with the Sgr dwarf galaxy punches a hole in the MW stellar disc that later winds up as a spiral with gaps \citep{BH21}, but dismiss this scenario, since any passage of Sgr dwarf was far outside the location of the gap. In our scenario, however, such a wide interaction is welcome, as it may trigger a grand-design spiral in the gas disc well inside the actual passage, such as for M51. In fact the simulated galaxy underwent two similar interactions (objects of $\sim10^8\Msun$ passing within $\sim20\,$kpc), which may be responsible for the grand-design spiral pattern that later caused the angular-momentum gaps in the young stars.

Another obvious explanation for any gaps in phase space is some orbital resonance with a regular non-axisymmetric perturber, in particular the Galactic bar. A prediction of non-linear perturbation theory is that a bar (or stationary spiral pattern) may drive a warp outside its outer 2:1 (Lindblad) resonance (OLR; \citealt{MassetTagger1997}). While the angular-momentum gap at $R\approx13\,$kpc indeed coincides with the steepening of the Galactic warp, it is outside the OLR of the bar. \citetalias{D22} therefore considered the outer 1:1 resonance with the bar, which is expected at that location, but could not find convincing evidence for an associated feature in any other tracer, including millions of stars from the Gaia radial-velocity sample and an RGB sample (both with 6D phase-space information). This essentially rules out such an explanation, since resonance mechanisms differentiate stellar populations based on their velocity dispersion, and hence should affect all disc populations, albeit more so those with lower velocity dispersion. However, this remains unexplored for evolving resonances, in particular a slowing bar which drives some age discrimination \citep{Chiba2021}. 

Our proposed scenario, on the other hand, differentiates directly on stellar age, strongly favouring young stars, since (i) it requires a special configuration of their birth places, which may not exist for extended periods of time, and (ii) the feature gets quickly erased by phase-mixing and angular-momentum dispersal due to spiral and bar perturbations. In the simulated galaxy, for example, we find that gaps in the angular-momentum distributions can be easily detectable only for stellar particles younger than about 1\,Gyr.

It has been widely studied that the interaction between the MW and the Sgr dwarf galaxy could have created a warp (e.g. \citealt{Ibata98}), a corrugation (e.g. \citealt{Gomez13, Laporte18}), the phase-space spiral (e.g. \citealt{Binney18,Laporte19}) and spiral arms (e.g. \citealt{Purcell11}) in the MW stellar disc. It would have been surprising if only the vertical structure of stellar and gas disc are affected by such an interaction, and the horizontal structure remained unaffected. Unfortunately, owing to the difficulty of obtaining the distance of \ion{H}{I} emitting gas, observational evidence for the MW gaseous spiral arms is quite limited \citep{Koo2017}. An alternative, albeit possibly distorted, snapshot of the gas disc is provided by the young stars like CCs.

\section{Summary}
\label{sec:sum}
We present an explanation for the gap at $L_\mathrm{gap}=2950\,\kms$kpc, corresponding to $R=13\,$kpc, found in the distribution of the angular-momenta $L$ of Milky-Way (MW) classical Cepheids (CCs; \citetalias{D22}, see Fig.~\ref{fig:gap1}) and illustrate it with a simulated galaxy from IllustrisTNG run TNG50-1. The simulated galaxy has an extended spiral structure which is traced by both the gas and the stars corresponding in age to the MW CCs. The distributions of these young star particles over radius, tangential velocity, and angular momentum resemble those for the MW CCs. In particular, their distributions over $L$ is multi-modal in all quadrants with at least one clear gap, caused by a spatial gap between their birth places in two spiral arms at the same azimuth. While the precise formation mechanism for these spirals cannot be extracted from the TNG50 run (owing to the low output cadence), the recent close flybys of two $\sim10^8\,\Msun$ objects within $\sim20\,$kpc of the simulated galaxy is likely to have played a role.

For this same mechanism to be the origin of the gap found in the MW CCs, two spiral arms on either side of $R=13\,$kpc must have formed stars 100-200\,Myr ago. While tracing the spiral structure of the Galactic ISM is notoriously difficult, the MW is not short of massive perturbers, such as the Sgr dwarf galaxy, which could have triggered or enhanced the spiral structure in the outer Galactic disc, leading to the formation of two spiral arms at the same azimuth and in the required radial range.

This explanation makes two predictions. First, that the angular-momentum gap at this position is \emph{only} present in populations of young stars, whose phase-space distribution still reflects their birth places. For older stars, phase-mixing together with bar and spiral perturbations quickly erases any coherence from a common origin. The precise age beyond which this feature should disappear is hard to predict, but based on the simulated galaxy we expect it to be around 1\,Gyr. The absence of a similar and clear phase-space feature in the general stellar disc (\citetalias{D22}), therefore corroborates our explanation and largely excludes a purely dynamical origin, such as orbital resonances with the Galactic bar, which affects all disc stars to some degree, depending on their velocity dispersion.

The second prediction is that the gap may coincide with some discontinuity in the properties of the warp, because two spirals in different parts of the warped gas discs, will have distinct orbital planes. This is the case in the simulated galaxy, which has a strong warp, but also for the MW CCs, where the gap visibly separates the stars in two populations in $z$.

\section*{Acknowledgements}
We are grateful to R.~Drimmel and S.~Khanna for providing the ages and distances of classical Cepheids from \citetalias{Gaia22} and to the IllustrisTNG team for making their simulations publicly available. We appreciate insightful discussions with Hossam~Aly and Ewa~L.~{\L}okas. This work was supported by STFC grant ST/S000453/1. MS thanks B.-E.\ Semczuk for support. EA thanks the CNES for financial support. RS thanks the Royal Society for generous support via a University Research Fellowship.

\begin{figure*}
	\includegraphics[height=7.5cm]{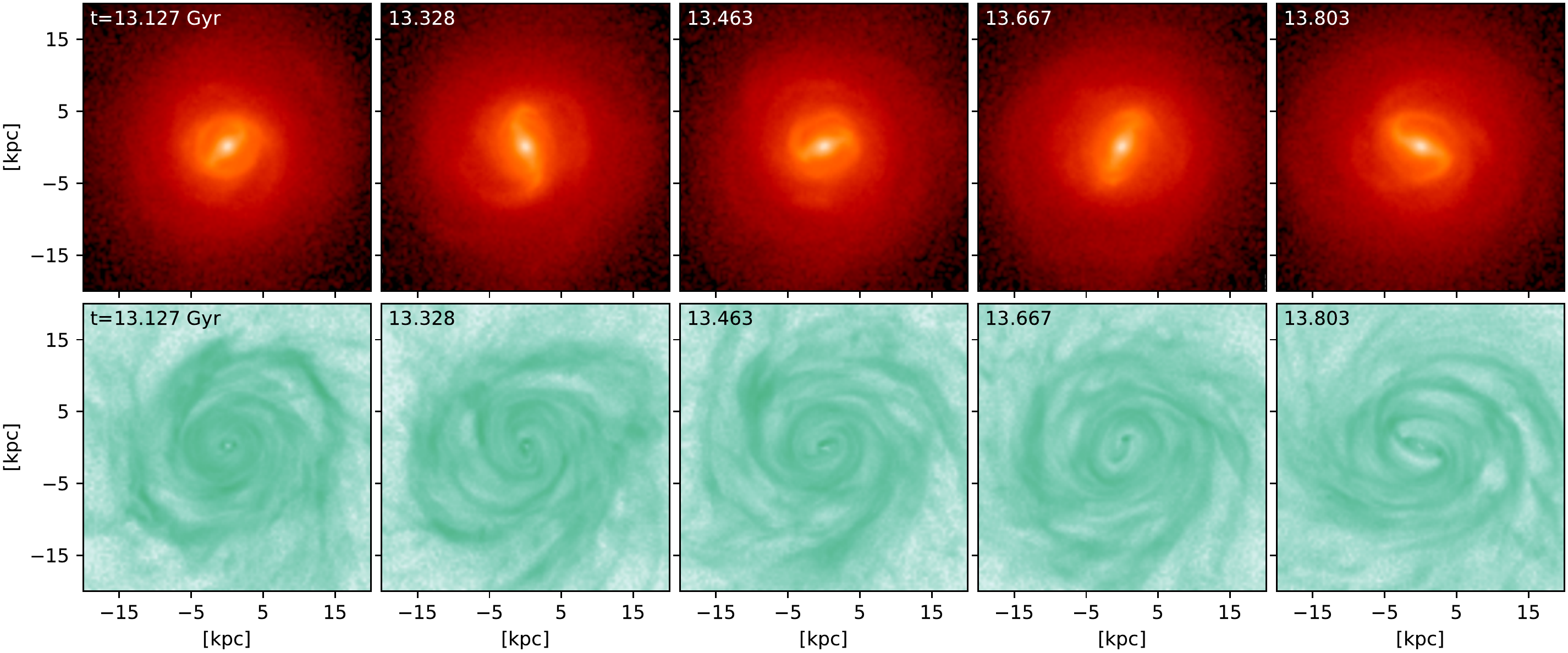}
    \vspace{-4mm}
    \caption{Face-on surface density distribution of all star (top) and gas (bottom) particles of the simulated galaxy for the last five snapshots, demonstrating the recent evolution of spiral structure.}
    \label{fig:map_evo}
\end{figure*}

\section*{Data Availability}
No data were generated in this study.



\bibliographystyle{mnras}
\bibliography{example} 





\appendix

\section{Evolution of the simulated galaxy}

Over the past 3-4 Gyr the simulated galaxy experienced several mild tidal interactions from flybys of smaller objects. At $t=12.8\,$Gyr, the time of bar formation, for example, two sub-haloes with total masses of $10^{9}\,\Msun$ and $5\times10^{8}\,\Msun$, respectively, passed within about $30$\,kpc and $20$\,kpc from the galactic centre. These interactions might have induced the bar and $m=2$ spirals (we cannot ascertain this from the low-cadence output of TNG50). Fig.~\ref{fig:map_evo} presents the snapshots of the face-on surface density of the stellar and gaseous discs in the last 680\,Myr, showing vivid evolution of the $m=2$ stellar spiral, whose asymmetric nature (e.g.\ at $t=13.328\,$Gyr) suggests a tidal origin. The grand-design spiral in the last snapshot, which we analysed in detail, might have been induced by the close ($\sim20$\,kpc) preceding passages of two light sub-haloes of $1.2\times10^{8}\,\Msun$ and $0.4\times10^{8}\,\Msun$, respectively. While their masses are below the limits for spiral-arms induction suggested by idealised simulations \citep{Pettitt2016}, \textsc{subfind} might slightly underestimate these masses of objects already embedded in the main halo. The near-simultaneous passages of two separate sub-haloes can interfere in constructive (or destructive) ways. Owing to the presence of mild interactions and other processes that could possible affect the morphology, such as misaligned gas accretion, we cannot determine the precise formation mechanism of the spirals with any certainty without re-simulating  the past 2\,Gyr in isolation.
Nevertheless, it seems likely that the Milky Way exhibited stronger tides from its satellites. For Sagittarius, for example, the initial mass ranges from $0.9-3.8\times10^9\,\Msun$ with last pericenters $\sim20$\,kpc varying between different models (e.g. \citealt{AdP21,Vasiliev21,Wang22}).

\bsp	
\label{lastpage}
\end{document}